Toward a Phenomenology of Computational Thinking in STEM Education

Pratim Sengupta

University of Calgary

Amanda Dickes

Harvard University

Amy Farris

Vanderbilt University

Author Note

Correspondence can be sent to: Pratim.sengupta@ucalgary.ca. Partial support from NSF CAREER Award #115230 and the Imperial Oil Foundation is gratefully acknowledged. All opinions are the author's and not endorsed by funding agencies.

Abstract

In this chapter, we argue for an epistemological shift from viewing coding and computational thinking as mastery over computational logic and symbolic forms, to viewing them as a more complex form of *experience.* Rather than viewing computing as regurgitation and production of a set of axiomatic computational abstractions, we argue that computing and computational thinking, should be viewed as discursive, perspectival, material and embodied experiences, among others. These experiences include, but are not subsumed by, the use and production of computational abstractions. We illustrate what this paradigmatic shift toward a more phenomenological account of computing can mean for teaching and learning STEM in K12 classrooms by presenting a critical review of the literature, as well as by presenting a review of several studies we have conducted in K12 educational settings grounded in this perspective. Our analysis reveals several phenomenological approaches that can be useful for framing computational thinking in K12 STEM classrooms.

*Keywords*: Computational thinking, Integrated STEM, modeling, phenomenology, science education

Toward a Phenomenology of Computational Thinking in STEM Education

## Introduction

In this chapter, we argue for an epistemological shift from viewing coding and computational thinking as mastery over computational logic and symbolic forms, to viewing them as a more complex form of *experience*. Rather than viewing computing as regurgitation and production of a set of axiomatic computational abstractions, we argue that computing and computational thinking, should be viewed as discursive, perspectival, material and embodied experiences, among others. These experiences include, but are not subsumed by, the use and production of computational abstractions. We illustrate what this paradigmatic shift toward a more phenomenological account of computing can mean for teaching and learning STEM in K12 classrooms by presenting a critical review of the literature, as well as by presenting a review of several studies we have conducted in K12 educational settings grounded in this perspective.

Papert (1987) famously referred to *technocentrism* as the fallacy of referring all questions about technology to the technology itself. A critical look at the history of educational computing tells us that the research in this field has also been predominantly technocentric in nature. Calls for taking into account the learners' perspective certainly have been made (e.g., Papert, 1980; DiSessa, 2000; Guzdial, 2008). However, the predominant effect of this call has also been technocentric in the sense that it has resulted in the creation of a new genre of programming languages (e.g., LOGO, Scratch, NetLogo, StarLogo TNG, AgentSheets, ViMAP, CTSiM, etc.) and microcontrollers (e.g., Arduino) designed to be easily usable for the "novice programmer". The technocentric focus is also evident in the learning objectives and assessment of computational thinking, which predominantly focus on the production and use of computational abstractions (e.g., see the studies reviewed by Grover & Pea (2013a)). Only a few, recent examples have focused on phenomenological aspects of computational thinking, such as the centrality of discourse (Grover & Pea, 2013b; Farris & Sengupta, 2014), the role of embodied reasoning (Francis, Khan & Davis, 2016) and the importance of managing, rather than ignoring uncertainty (Farris, Dickes & Sengupta, 2016) in the development of computational thinking in STEM curricular contexts. And, while recent arguments have been made for an increased awareness for paying attention to sociological dimensions of computing such as virtual communities (e.g., online SCRATCH communities) and out-of-school, DIY makerspaces (Kafai & Burke, 2013), our focus here is on the K-12 public school classroom.

Our chapter is an argument for deepening and broadening the focus on the phenomenology of computing and computational thinking in K12 STEM curricular contexts and classrooms. Our concerns are both epistemological and pedagogical, and are grounded historically as well as in the pragmatics of K12 classrooms with the focus on sustaining computing as a long-term practice. The first part of the chapter presents a critical and synthetic review of the literature and argues for a phenomenological epistemology of computational thinking that foregrounds the uncertainty and complexity in the experience of computing and science, in professional practice and in STEM classrooms. The second part of the chapter presents a set of pedagogical approaches for sustaining computing and computational thinking through computational modeling in the STEM classroom. This is presented in the form of a critical review of studies that conducted by our research group in K12 classrooms in the US, including studies that were conducted in the form of partnerships with teachers.

### Phenomenology of Computational Thinking

Since the phrase "computational thinking" has been popularized by Wing (2006), there have been a plethora of studies on computational thinking in education. Yet, beyond the early work on computational literacy by Papert (1980) and diSessa (2001), the *epistemology* of computational thinking has received very little attention in the literature. In this section, we examine core beliefs and assumptions about the nature of knowledge and knowing that are and should be involved in thinking computationally, by adopting a historical perspective as well as reviewing recent advances in educational computing, from a phenomenological perspective. We highlight the importance of grounding computational thinking in representational and epistemic practices that are central to *knowing* and *doing* in science, and more broadly, in STEM education. We also argue that thinking carefully in terms of these practices can help us understand the uncertainty and subjectivity inherent in the experience of computational thinking in STEM, which is essential for sustainable and long-term curricular integration of computational thinking.

**Inseparability of Abstractions and Practices in Computing and Science**

Wing (2006) defined the phrase "computational thinking" to indicate a "thought process involved in formulating problems and their solutions so that the solutions are represented in a form that can be effectively carried out by an information-processing agent" (Wing 2006, p 1). According to her, the "nuts and bolts" in computational thinking involve dealing with abstractions in the following ways: a) defining abstractions, b) working with multiple layers of abstraction, and c) understanding the relationships among the different layers (Wing 2008). Abstractions, according to Wing (2006), give computer scientists the power to scale and deal with complexity. She noted:

> Abstraction is used in defining patterns, generalizing from instances, and parameterization. It is used to let one object stand for many. It is used to capture essential properties common to a set of objects while hiding irrelevant distinctions among them. (Wing 2006, p1).

Wing's conceptualization of abstraction, as the excerpt above shows, therefore, emphasizes the notion of generalization. Abstractions, in her view, are generalized computational representations that can be used (i.e., applied) in multiple situations or contexts. In this sense, as Sengupta et al. (2013) pointed out, her definition of abstraction is similar to Locke's. In Locke's view, abstraction is the process in which "ideas taken from particular beings become general representatives of all of the same kind" (Locke 1690/1979).

However, a phenomenological interpretation of Wing's notion of abstractions is incomplete without a deeper understanding of the contextualization that necessitates and ground computational abstractions in professional practice. For example, Schmidt (2006) points out that software researchers and developers typically engage in creating abstractions that help them program in terms of their contextualized design goals - e.g., the specific problem that they are solving, which is often in a different field (domain) of professional practice. The abstractions that "need" to be created are essential because the end-user must be shielded from avoidable complexities, such as the CPU, memory, and network devices, and instead, interact directly with the domain-specific problem (Schmidt, 2006).

In this perspective, the term "thinking" in computational thinking is perhaps a semantic reduction of its intended meaning, because phenomenologically it involves both representational and epistemic work. Sengupta et al. (2013) therefore argued when the notion of computational abstractions is grounded *in use*, it could be understood as a practice that draws upon concepts that are fundamental to computing and computer science, and, it also includes practices such as problem representation, abstraction, decomposition, simulation, verification, and prediction that are also central to modeling, reasoning and problem solving in a large number of scientific, engineering and mathematical disciplines (National Research Council, 2008; NGSS, 2015).

Sociologists and philosophers of science have also identified the inseparability of abstractions and practice in the work of scientists. It is rarely the case that the transformation of an initial idea to a successful scientific experiment or a model is a simple and linear process that relies on solely the invention and use of abstractions. The philosopher Andrew Pickering pointed out that scientists are always enmeshed in a "mangle of practice" (Pickering, 1995). That is, scientists struggle continuously in order to get theories and instruments on one hand and the natural world on the other to perform in the ways that their investigations require. The creation of scientific knowledge can therefore be understood as a dynamical process of interactive stabilization of material and human agency (Pickering, 1995; Lehrer, 2009). Uncertainty, and managing uncertainty are unavoidable aspects in this work, even though the most popular image of scientific work tends to be one of the certitude of accurate predictions (Duschl, 2009).

A central focus of the scientific work is the invention, re-production and modification of scientific inscriptions – such as graphs, equations, computer code, etc. – which tend to amplify certain aspects of the phenomena under investigation while reducing emphasis on other, less relevant aspects (Latour, 1990). This is similar and synergistic to the work of defining and using contextually relevant computational abstractions, as we pointed out earlier. Additionally, computational models can also bring to light new, unexpected ways of thinking about the phenomena by bringing different disciplinary perspectives in contact with one another (McLeod & Nercessian, 2015). The process of creation of these inscriptions – that are collectively termed "modeling" - involves both representational and epistemic work in a deeply intertwined manner (Giere, 1988; Nercessian, 1996; Pickering, 1995; Lehrer, 2009). This perspective is known as the "science as practice" perspective, and is now regarded as a cornerstone of science education research (NGSS, 2015). In the following sub-sections, we consider the subjective and perspectival nature of the work involved in modeling, in particular, computational modeling.

**Subjectivity in Representational Work**

An equation is a model, so is a simulation. Studies of scientists and the production of scientific inscriptions, during various stages of the production and refinement of scientific inscriptions reveal a rather amorphous nature of scientific knowledge and work (Pickering, 1995; Ochs, Gonzales & Jacoby, 1996; Latour, 1996; Daston and Galison, 2007). For example, Ochs, Gonzales & Jacoby (1996) highlighted the central role that interpretive work, including negotiation between scientists play in dealing with uncertainty during a research project. They also demonstrated that the interpretive nature and uncertainty of this work – an epistemic phenomenon – is deeply tied to the representational infrastructure (Ochs, Gonzales & Jacoby, 1996). This is echoed by other scholars such as Daston and Galison (2007), who pointed out that as representational technologies evolve and new representational technologies emerge, they necessitate new forms of uncertainty and interpretive work.

Daston and Galison's (2007) argued that with the introduction of photographic technology and the printing press, the epistemic stance of scientific work shifted from a falsely "objectivitist" stance to "trained judgment". This was evident in their comparison between the 19th century introduction of photographic technology where the machinic nature of photography created an impression that scientist could "get out of the way," and let the photograph produce what became perceived as bare, un-interpreted, objective "facts." In contrast, beginning in the early to mid-twentieth century, with the advent of the printing press that in turn widened the audience for scientific works such as atlases, the production of scientific images became necessarily more interpretive on part of the scientist, with a clear goal of *enhancing the communicativity* of the images, which Daston & Galison (2007) termed "trained judgement".

Building on this work, Farris, Dickes & Sengupta (2016) have argued that the advent of computing as a *key* mode and medium of scientific inquiry further amplifies this epistemic stance of "trained judgement". A case to point, they argued, is that recent, long-term ethnographic studies of biomedical engineering labs illustrate how the malleability and inherent interdisciplinary of the practice of computational modeling results in new conceptual innovations in scientific practice (Nersessian, 2012; Chandrasekharan and Nersessian, 2014, 2015). Nercessian and colleagues showed that computational modeling can be particularly helpful for creating new scientific knowledge in the field of complex systems, by: a) bridging the gap between theorization, dynamic visualization and experimental work, b) bringing together multiple disciplinary perspectives; c) use stochastic modeling techniques in cases where clear mechanistic accounts are difficult to obtain; and d) making it possible to communicate directly with colleagues about complex, predictive visualizations of the target phenomena.

**Computational Modeling as Perspectival Work**

In his seminal book *Mindstorms*, Papert argued that working with the LOGO turtle is a "model for what it is to get to know an idea the way you get to know a person" (Papert, 1980, pp 136). Papert argued that it involves *getting to know* the turtle, through exploring what it can or cannot do. He cautioned that this should not mean that all ideas be reduced to computational terms; rather, the early experience with turtles is a good model of learning. That is, "… it is a good way to "get to know" subject by "getting to know" its powerful ideas" (Papert, 1980, p 138). As an illustrative case, he noted that when children learn Newtonian mechanics using LOGO, they do so through modeling changing velocities, i.e., by specifying how fast the turtle should move. The propositional forms of these phenomena are represented in the form of physical laws in the form of linear mathematical equations, and the fallacy of education is that these laws, which are the products of complex work (i.e., Pickering's mangle of practice) in which qualitative thinking that is less completely specified and seldom stated in propositional form play an important role. Therefore, it is the qualitative experience of *thinking like the turtle* and *thinking with the turtle* that makes the experience of learning a powerful and a deep one, and one that is quite antithetical to *learning as usual* in K-12 science (and beyond). These forms of reasoning enable the learner to engage in embodied and intuitive reasoning (Papert, 1980; Wilensky & Resiman, 2006; Sengupta & Wilensky, 2009).

The early success of LOGO has led to the development of several LOGO-like programming languages and modeling environments such as NetLogo (Wilensky, 1999), Scratch (Resnick et al., 2009), AgentSheets (Repenning, 1999), CTSiM (Sengupta et al., 2013; Basu & Biswas, 2016), ViMAP (Sengupta et al., 2015b). Computational models developed in such languages are more generally known as agent-based models (ABMs). When users develop ABMs, they construct programs by providing simple rules to a computational object or agent such (e.g., the sprite in SCRATCH, the turtle in LOGO, etc.), which then enacts the rules through movement in computational space. These agent-level actions are repeated over time, and/or across multiple agents. In the former case, it enables learners to generate models of continuous movement (Newtonian mechanics) from temporal aggregations of discrete actions (Sengupta & Farris, 2012; Sengupta, Farris & Wright, 2012). In the latter case, it enables learners to model dynamical systems (e.g., ecological interdependence) in which multiple agents are simultaneously interacting with each other (Dickes & Sengupta, 2013; Dickes et al., 2016).

Because the agent-level interactions, attributes and behaviors are often *body-syntonic* (i.e., can be explained and understood through simple embodied actions of the child), young children can model complex scientific phenomena using such forms of computing (Papert, 1980; Danish, 2014; Dickes et al., 2016; Levy & Wilensky, 2008). As Dickes et al. (2016) demonstrated, by engaging in agent-based modeling, even young learners can investigate and develop explanations of system-level, emergent behaviors from the perspective of agents within the system. They key argument supported by these studies is that thinking like the agent provides learners an intuitive pathway in exploring emergent outcomes of the system (Wilensky & Reisman, 2006; Levy & Wilensky, 2008). Evelyn Fox Keller's biography of the biologist Barbara McClintock supports this claim, citing evidence that thinking like the agent (e.g., a chromosome) enabled McClintock to make significant advances in her research on human genetic structures (Keller, 1983). Similarly, Ochs, Gonzales, and Jacoby (1996) also identified that scientists' sensemaking in the domain of physical sciences also involves such mental projections of the self into the phenomenon of inquiry.

### Phenomenological Approaches for Sustaining Computing in STEM Classrooms

What does the theoretical review in the preceding section mean for the praxis of computing in STEM education? We argue that the *experience* of coding in STEM, from the perspective of the learners and teachers, especially over a long period of time, is inherently heterogeneous. That is, dealing with computational abstractions in the context of STEM disciplinary contexts and classrooms involves engaging with multiple forms and genres of representations beyond coding, and often translating between these representations require interpretive judgements. This stands in contrast to the views that have been more traditionally supported by educational researchers, where the goal is to "apply" algorithmic thinking and computational abstractions to determine the correct answer. This complexity is left out in techno-centric images of coding, even when they apparently focus on computational productions by participants.

In the remainder of this section, we propose some phenomenological approaches that can help us address these issues in the K-12 STEM classroom. We will review a set of studies conducted in partnership with K-12 teachers and students. Participants in these studies used coding in order to design and develop models in science and math on a long-term basis, throughout the academic year. We present a close examination of the nature of the experience through which teachers and students appropriated coding and computational thinking as the language of doing scientific work in their classrooms. We begin with an argument for adopting a particular *genre* of programming and modeling  (*agent-based* programming and modeling) for modeling across disciplines, which is essential for long-term curricular integration. We then suggest a set of pedagogical guidelines for integrating programming in the K-12 STEM curricula, grounded in the perspectives of teachers and learners in K-12 classrooms.

**Agent-based Computational Modeling as a Trans-Disciplinary Practice**

Scientific practices like modeling develop only over the long term, both historically within the sciences and ontogenetically, within the lifetime of individuals. This is because modeling is a rather nuanced and complex form of epistemology, even though most educational texts and curricula do not directly address these complexities (Lehrer, 2009). The *year-long science classroom* is a better context for engaging children in such extended forms of practice, rather than the predominant tradition in educational research to conduct intervention studies where children engage in modeling (including computational modeling) spanning a few hours to a few days. But, in order to support such long-term curricular integration, we must take into consideration how to integrate computational modeling and programming *across* disciplinary contexts.

Different forms of phenomena lend themselves to different forms of modeling (Lehrer & Schauble, 2007), and we have found that at the elementary, middle and high school levels, the categories of *linear continuity* and *emergent aggregation* can be helpful guides for us in selecting scientific phenomena across disciplines that can lend themselves well to computational modeling and programming. An example of modeling *linear continuity* would be modeling motion as a continuous change in position, where the behavior of a single "agent" (e.g., a ball rolling on a ramp) can be modeled as a temporal series of changes of position and/or other variables such as speed and acceleration that obey linear mathematical relationships (Sherin et al., 1993; Sengupta & Farris, 2012). An example of modeling *emergent aggregation* would be modeling ecological interdependence, where multiple agents simultaneously interact with each other and the environment, which in turn result in aggregate level outcomes, e.g., the dynamical relationship between the predator and prey populations in an ecosystem (Wilensky & Reisman, 2006; Dickes & Sengupta, 2013; Wagh, Cook-Whitt & Wilensky, 2017). Such aggregate level behaviors or outcomes are known as *emergent*, because although linear relationships between individual agents (objects) produce these behaviors, these behaviors are not apparent in the description of either the individual objects or the relationships (Lehrer & Schauble, 2007; Wilensky & Resnick, 1999). Other examples of emergent phenomena that have been successfully adopted by teachers and students through the use of agent-based computational modeling and programming include electrical conduction (Sengupta & Wilensky, 2011), crystallization (Blisktein & Wilensky, 2009), molecular chemistry (Stieff & Wilensky, 2003; Levy & Wilensky, 2012), evolution (Novak & Wilensky, 2010), ethnocentrism (Hostetler, Sengupta & Hollet, in press), etc. This suggests that adopting agent-based modeling and programming as the form of computing can make it possible for educators to use the same genre of modeling and programming across multiple disciplines.

It has also been argued that students' conceptual difficulties in understanding both linear continuity and both emergent aggregation have similar origins (Reiner, Slotta, Chi & Resnick, 2000; Chi, 2005). For example, Reiner *et al.* (2000) argued that physics novices tend to use substance-based knowledge when reasoning about concepts like force, heat, light, and electric current (*e.g.*, force as a property of an object). For example, the misconception that continuing motion implies a continued force in the direction of the movement (Clement, 1982) is generated from a more primitive idea (called phenomenological primitives or p-prims) called "continuous force," which can be abstracted from common everyday experiences of needing constant effort to keep an object in motion (diSessa, 1993). Note that these novice intuitive ideas about physics have an underlying structure of a direct schema -- one that involves an agent either acting on another agent, or, an agent being acted upon by an impetus (Talmy, 1983). On the other hand, an expert-like understanding of kinematics involves being able to conceptualize a situation in terms of more complex interactions – *e.g.*, situations involving lack of motion, or constant speed, according to an expert, are both forms of dynamic equilibrium between interacting systems (Clement, 1990; Greeno & Van de Sande, 2007). Similarly, in the domain of ecology, researchers have argued that commonly noted misconceptions are indicative of direct schema or event schema, that imply a direct cause effect relationship (such as "A" causes "B") or an event that has a finite duration of time (as opposed to being continuous), whereas the expert conception of ecological phenomena involve a more complex cognitive structure involving the dynamic and decentralized nature of emergent phenomena in terms of a myriad of simultaneous interactions (Chi, 2005). However, studies have also shown that pedagogical approaches based on agent-based models and modeling can act as productive learning environments, using which novice learners can develop a deep understanding of dynamic, aggregate-level phenomena by bootstrapping, rather than discarding their agent-level intuitions (Dickes & Sengupta, 2013; Dickes et al., 2016; Wilensky & Reisman, 2006; Levy & Wilensky, 2008).

This body of research also provides useful guidelines for the sequence of learning activities in each domain, and our general pedagogical approach explicitly adopts the perspective that expert-like scientific knowledge can result through building upon and refining existing naive intuitive knowledge (Dickes et al., 2016; Danish, 2014; Sengupta et al., 2015b). For example, the initial learning activities leverage a naive conceptualization of the domains, and progressively scaffold them towards refinement. In kinematics, learners begin by inventing representations of motion in terms of measures of speed (how fast an object is moving) and inertia (innate tendency of an object to continue its current state of rest or motion, which often takes an anthropomorphic form in novice reasoning), and gradually move to a force-based, more canonical description of motion in subsequent activities (Sengupta & Farris, 2012; Farris, Dickes & Sengupta, 2016). Similarly, in ecology, students begin with programming the behavior of single agents in the ecosystem and gradually develop more complex programs for modeling the behavior and interaction of multiple species within the ecosystem (Wilensky & Reisman, 2006; Danish, 2014; Sengupta et al., 2013; Dickes et al., 2015).

## Framing Programming as Designing Mathematical Measures of Change

Our studies have demonstrated that framing programming as "mathematizing" in the science classroom can serve as a productive pedagogical approach for integrating programming in the K12 science classroom (Sengupta et al., 2015; Sengupta et al., 2013; Dickes, Farris & sengupta, 2016; Farris, Dickes & Sengupta, 2016). In this approach, programming is used in the context of creating computational models of scientific phenomena through designing discrete mathematical representations of units of change, for representing change over time. That is, the computational code created by students serve to define a "unit" of measurement, which would then get repeated as the program was "run" to produce the desired motion.

From the perspective of praxis in the K-12 science classroom in North American public schools, this form of activity is of critical importance for classroom integration of computational modeling and programming. Teachers in US and Canadian public schools who we have worked with, have reported that interpreting and constructing mathematical measures (for example, units of measurement and graphs) is an area where most of their students experience difficulties. This is also of importance for US and Canadian public schools because manipulating units is emphasized in standardized assessments (in the US) and the program of studies (in Canada), and therefore, teachers acknowledge this as an important learning goal in their regular science classroom.

We see this as a great opportunity for integration of computational modeling and programming in K12 science classrooms. Our studies how that agent-based programming and modeling can help students overcome conceptual challenges in understanding linear continuity (e.g., kinematics, see Sengupta & Farris, 2012) and emergent aggregation (e.g., ecology, see Dickes et al., 2016a, 2016b), through the iterative design of measures of change over time. This is because the activity of programming the behavior of agents requires the learners to define the event in discrete measures (Sengupta et al., 2015b). The state of the simulation, at any instant, represents a single event in the form of spatialized representations of agent actions and interactions. To "run" the simulation, these events are repeated a number of times specified by the user. By engaging in iterative cycles of building, sharing, refining, and verifying computational models, students refine their understanding of what actions and interactions of agents represent an "event," which are then displayed on graphs. This enables students to define and explore different kinds of units, and see their simulation measured in those units (Farris, Dickes & Sengupta, 2016) and even merge computational modeling with artistic design (Sengupta, Farris & Wright, 2012).

**Supporting Perspectival Work Through Embodied Modeling**

Research in science education suggests that the integration of ABMs in elementary classrooms also benefits greatly from the use of other synergistic forms of modeling such as embodied and physical modeling. Programming an agent involves learning to think like the agent, because it can help students understand the relationship between their code and the simulated output. In our studies, all teachers saw embodied modeling as a valuable activity for teaching students how to think like an agent. Embodied modeling introduces the students to the relevant computational rules represented by the agent-based programming commands, helps them debug their programs and deepens their understanding of the graphs in the simulations (Dickes, Farris & Sengupta, 2016; Dickes, Sengupta, Farris & Basu, 2016).

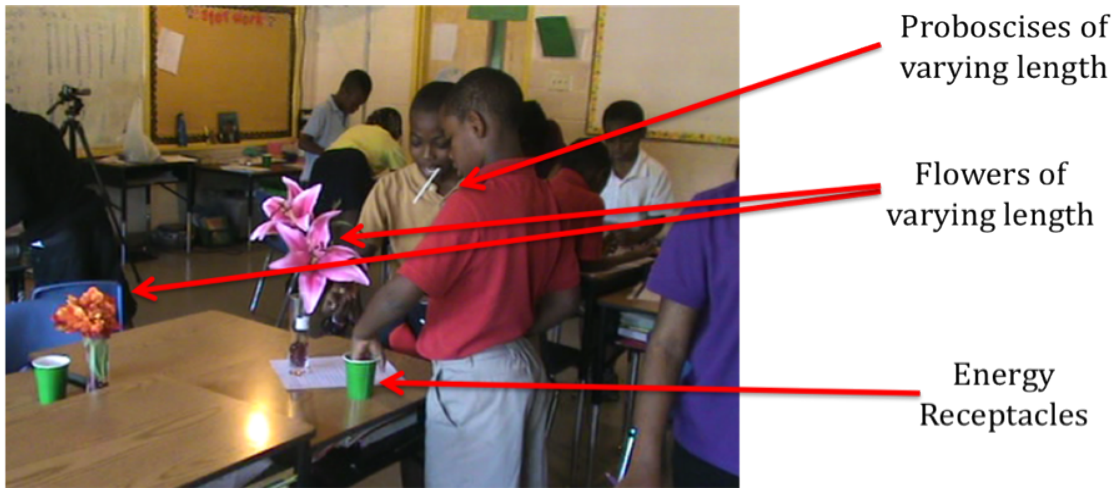

*Figure 1*: Students participating in Phase I's Embodied Modeling Activity

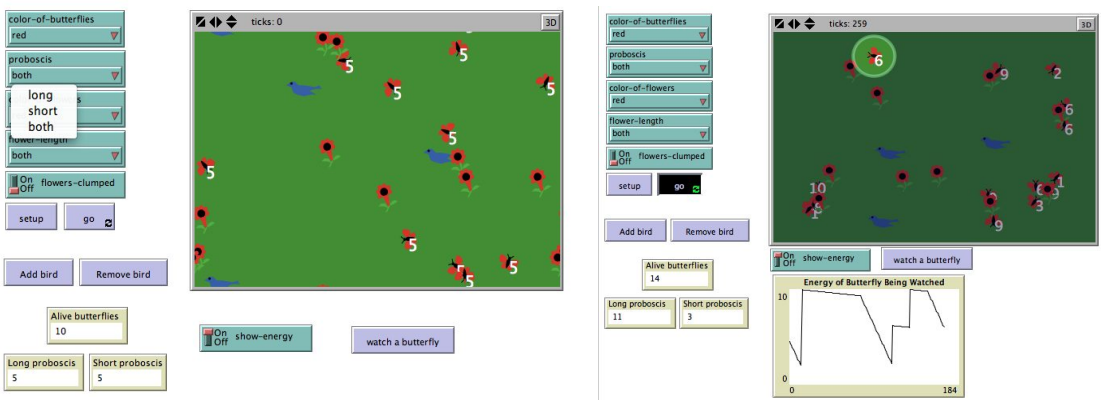

*Figure 2:* Screenshots of the Predator ABM (left) and Watched Energy ABM (right). Both models were designed to actively recruit students' previous embodied modeling experiences shown in Figure 1.

But why are these different forms of modeling necessary? Science educators and cognitive scientists have argued that embodied thinking is central to the development of agent-based thinking and representational practices (Papert, 1980; Goldstone & Wilensky, 2009;

Wilensky & Reisman, 2006). For example, in a recent study conducted in a 3rd grade classroom, students began with an embodied modeling activity of foraging behavior, followed with the generation of mathematical inscriptions based on their embodied actions, and finally, conducted further inquiry of interdependence in an ecosystem using two separate ABMs (Dickes et al., 2016). We found that the students recalled and built upon their embodied modeling experiences as butterflies foraging for nectar (see Figure 1), during their subsequent interactions with the agent-based simulation of a butterfly-bird-flower ecosystem (see Figure 2). We also found that creating mathematical inscriptions (bar graphs) to represent the data collected during the embodied modeling activity provided a representational continuity between the embodied modeling activities and the ABMs, as well as with previous representational forms that students used and developed in their science and math classes prior to the study. We also found that embodied modeling activities, especially in the case of modeling interactions between different types of agents, must be designed so that students are able to take on the perspectives of different types of agents, rather than prompting students to take on the perspective of only one type of agent.

As students engaged iteratively in cycles of embodied modeling and graphing, and then modeled the same phenomena using multi-agent based NetLogo simulations, we found that they were able to develop progressively more complex forms of mechanistic explanations of emergence. Mechanistic explanations focus on the processes that underlie cause–effect relationships and thereby take into account how the activities of the constituent components affect one another (Bolger, Kobiela, Weinberg, & Lehrer, 2012; Machamer et al., 2000; Russ et al., 2008). In particular, we found that  learners were able to engage in a particular form of mechanistic reasoning that Russ et al. (2008) termed  *chaining*. During chaining, learners use knowledge about the causal structure of the phenomena to make claims about what must have happened previously to bring about the current state of things (backward chaining ), or what will happen next given that certain entities or activities are present now (forward chaining ).

This is an important finding from the perspective of computational thinking in the context of science education, because this suggests that event-based programming and modeling can support children in developing deep conceptual understandings of complex scientific phenomena. Furthermore, this also suggests that focusing on supporting the growth of students' mechanistic reasoning through modeling may be helpful for integrating computational thinking in science classrooms. As Sengupta et al. (2013) identified, mechanistic reasoning in the domain of science education is well aligned with *algorithm design* and *complexity analysis* in the domain of computational thinking.

### Refining Computational Modeling through Disciplinarily-grounded Classroom Norms

Our studies also illustrate that emphasizing mathematizing and measurement as key forms of learning activities can help teachers meaningfully integrate programming as a "language" of science, and further, that teachers can accomplish this by supporting the development of sociomathematical norms (Dickes, Farris & Sengupta, 2016). Science educators have shown that the iterative design of mathematical measures can result in deep conceptual growth of students in elementary science, especially when these activities are integrated throughout the curriculum over several months (Lehrer, 2009). Lehrer and colleagues have also shown that the question of what counts as a "good" model also needs to be normatively established in classroom instruction in order to deepen students' engagement with scientific modeling in elementary grades, and that these norms also follow similar shifts toward deeper disciplinary warrants over time (Lehrer & Schauble, 2006; Ford & Forman, 2006; Lehrer, Lucas

& Schauble, 2008). In such classrooms, mathematical modeling becomes as a meaning-making lens through which the natural world can be systematized and described (Lehrer, Schauble, Strom & Pligge, 2001).

The specific genre of classroom norms that we have found to be at work in our studies has been termed *sociomathematical norms* (McClain & Cobb, 2001; Yackel & Cobb, 1996; Cobb, Wood, Yackel, & McNeal, 1992). In a recent paper, we outlined demonstrated how the emphasis on developing and refining sociomathematical norms pertaining to the design of mathematical measures of motion can help teachers seamlessly integrate programming with science education in a 3ʳᵈ grade classroom, and how they are taken up in students' work. Sociomathematical norms differ from general social norms that constitute the classroom participation structure in that they concern the normative aspects of classroom actions and interactions that are specifically mathematical. These norms regulate classroom discourse and influence the learning opportunities that arise for both the students and the teacher. As in the work of Cobb and his colleagues (Cobb, Yackel, & Wood, 1989; Yackel, Cobb, & Wood, 1991), we also found that it was the classroom teacher who initiated and guided the development of these norms in order to foster and sustain classroom microcultures characterized by explanation, justification, and argumentation.

An important, and rather fundamental sociomathematical norm that began as the central guiding question posed by the teacher at the beginning of the class was "what counts as a *good* model". Similar to Yackel and Cobb (1996), we found that this norm typically originated as a socially defined norm, and shifted over time to a more sociomathematically defined norm. That is, students' initial warrants were decided on the basis on how many of their peers "liked" a particular model during class discussion and sharing of models, rather than thinking more deeply about how their ViMAP code represented the relevant phenomenon they were modeling. However, over time, these warrants became progressively more grounded in the mathematically warrants of how representative their code were of the relevant phenomena being modeled. The class jointly took up normative ways of thinking about, and representing motion (walking) through designing and refining *approximate* and *predictive* measures of change over time, using embodied modeling activities, drawings of their embodied modeling activities that represented "step-sizes", and their ViMAP code and graphs (Dickes, Farris & Sengupta, 2016).

Overall, we found that students' use of the ViMAP programming commands became increasingly sophisticated as they held their models accountable to the sociomathematical norms (Dickes, Farris & Sengupta, 2016).  Over a six-week period, we scored each student's final ViMAP model at the end of each class period in terms of whether they used appropriate computational abstractions identified by Sengupta et al.  (2013) as being relevant to computational thinking such as variables, loops and initialization.  Students' code were scored on the appropriate and non-redundant use of variables and loops in their models, and whether their graphs represented appropriate element(s) of the phenomenon being simulated using their ViMAP code. The growth in students' computational fluency is evident in Figure 3. For example, a score of zero meant none of the variables used were appropriate, whereas a score of 3 meant no use of redundant or incorrect variables. The accuracy of the graphs in students' later models were indicative of the appropriate use of the "repeat" command (i.e., loops), and order of placement of the "place measure" command. This in turn relied on their conceptual understanding of when to initialize the measurement (i.e., initialization), and how often the desired measurement had to be repeated in order to generate the graph (loops).

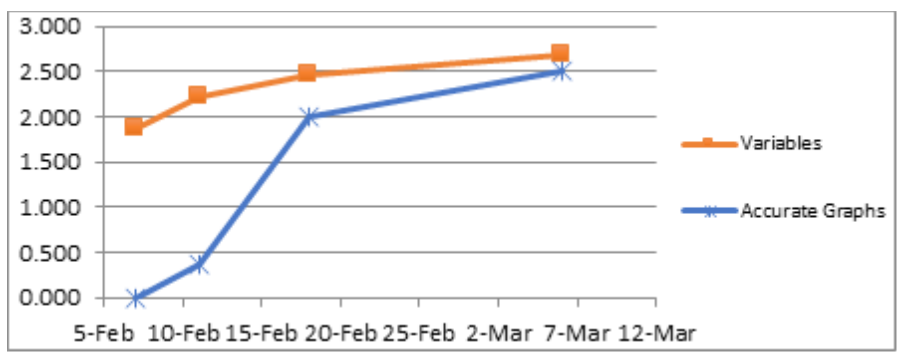

*Figure 3:* Improvement in Computational Thinking Supported by Socio-Mathematical Norms

**Framing Coding as Designing for Authentic Use**

In a study conducted in a 4th grade classroom in a low income (90% free lunch), public charter school in Nashville, we investigated how collaboratively designing computational machines for authentic users could support the integration of coding in STEM education (Sengupta et al., 2015b). The first phase of the study focused on introducing students to agent-based programming through creating geometric shapes (e.g. squares, circles, spirals) using ViMAP. This phase lasted for 8 class periods. For the next 18 class periods, students worked in dyads on a STEM design challenge (capstone activity), i.e., constructing mathematical machines and user guides for generating geometric shapes using a distributed computing infrastructure.

During the capstone learning activity in both the studies, learners worked in dyads and constructed a mathematical machine for generating geometric shapes. Each machine consists of two components: virtual and physical. The virtual component was a ViMAP program, that learners constructed using visual programming primitives selected from the ViMAP programming library. The physical component consisted of two physical control interfaces, each designed to control the reading on one of the distance sensors. Each sensor controlled a distinct Turtle-variable (e.g., color, speed, rotation). This was an activity that required intersubjective collaboration (Sengupta et al., 2015b), because while each member of the dyad independently designed one of these physical control structures using Lego bricks, the dyad was responsible for jointly designing the ViMAP program. Figure 4 shows an example of student work.

We specified that other 4th-grade teachers in Nashville would use these machines, so that students had a specific image of user(s) in mind. To ensure authenticity of the users, we also invited three graduate students in education with prior math teaching experience in elementary grades, but unaffiliated with our study, to serve as "users". The user testing took place twice: first in mid March (User Testing 1), and in late April (User Testing 2). During both the User Testing events, each user interacted with a dyad's machine for about 20 minutes, and provided them written and verbal feedback. After User Testing 1, students improved their machines and user guides in order to address the issues highlighted in the feedback. User Testing 2 was also the capstone activity.

We compared the work of each dyad at two stages: User Testing 1 (UT1), and User Testing 2 (UT2). In terms of children's mechanistic explanations, we found that compared to UT1, attending to what the user needs to know resulted in improving greatly the quality of students' mathematical explanations during UT2. Their explanations, as evident both in their user guides and verbal explanations during the user testing process, made explicit the mathematical relationships between algorithmic elements (e.g., number of loops in their ViMAP program) and variables in their ViMAP programs, and the actions of the Turtle in every step (e.g., right turn), which in turn was directly effected by the users' actions (e.g., sensor reading generated by the user).   The greater emphasis on identifying and representing the relationships between computational abstractions (algorithms and variables), mathematical relationships and the mechanics of the physical setup resulted from the need to create designs that were more *communicative* (Sengupta et al., 2015b). A sample comparison is shown in Figure 5.

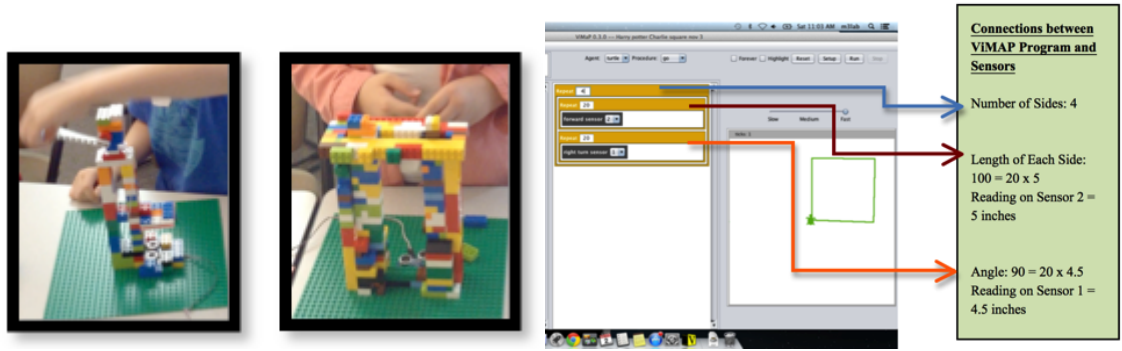

Figure 4A (Left): Jerry's pulley mechanism for controlling turn of the Turtle via Sensor 1; Figure

4B (Middle): Chuck's machine for controlling the speed of the Turtle via Sensor 2. Figure 4C

(Right) is a screenshot of their ViMAP program for generating a square, and our annotation makes explicit the multiplicative reasoning involved in generating *angles* and *sides* of the square.

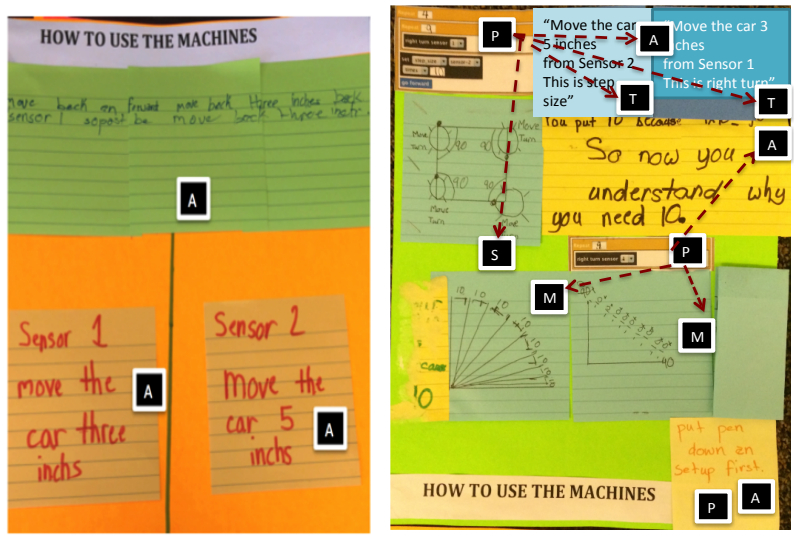

Figure 5a: Jacinda and Tom's user guides in User Testing 1 (left) and User Testing 2 (right). We annotated their user guides using the schematic shown in Figure 5b.

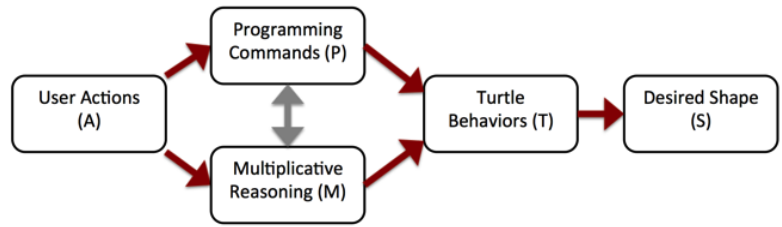

Figure 5b: A schematic for mechanistic explanations used by all groups in User Testing 2

The phenomenological lesson here is that when coding is embedded in an authentic design activity intended for, and tested by an authentic audience, paying attention to the needs and the perspective of the user can deepen the coders' conceptual understanding of the relationship of computational abstractions with disciplinarily grounded knowledge and representations.

**Support Transition from Visual to Text-based Programming**

Another important issue for sustaining programming in K12 STEM classrooms, especially in higher grades (middle school or high school), is that although visual programming lowers the barrier for entry into programming, learners who intend to pursue careers in computing may find the drag-and-drop nature of visual programming inauthentic, or find it difficult to transition to text-based programming (DiSalvo, 2014). In a recent study conducted in a 8[th] grade classroom, we investigated this issue (Sengupta et al., 2015b). We used ViMAP, because ViMAP is a dual-mode programming language that enables users to engage in both blocks and text-based programming. Visual programming commands in ViMAP are defined as short NetLogo procedures (see Figure 6), which students can easily access and modify using text-based NetLogo code. In our study, after engaging in visual programming with ViMAP for approximately two months to build simulations of interdependence in ant ecosystems, the teacher and the students wanted to make deeper changes in the underlying text-based NetLogo code. But, given the limited instructional time, the teacher found it challenging to help students create new simulations in NetLogo using text-based programming. This required a lot of "overhead," because the language syntax was often disconnected from the relevant scientific concepts.

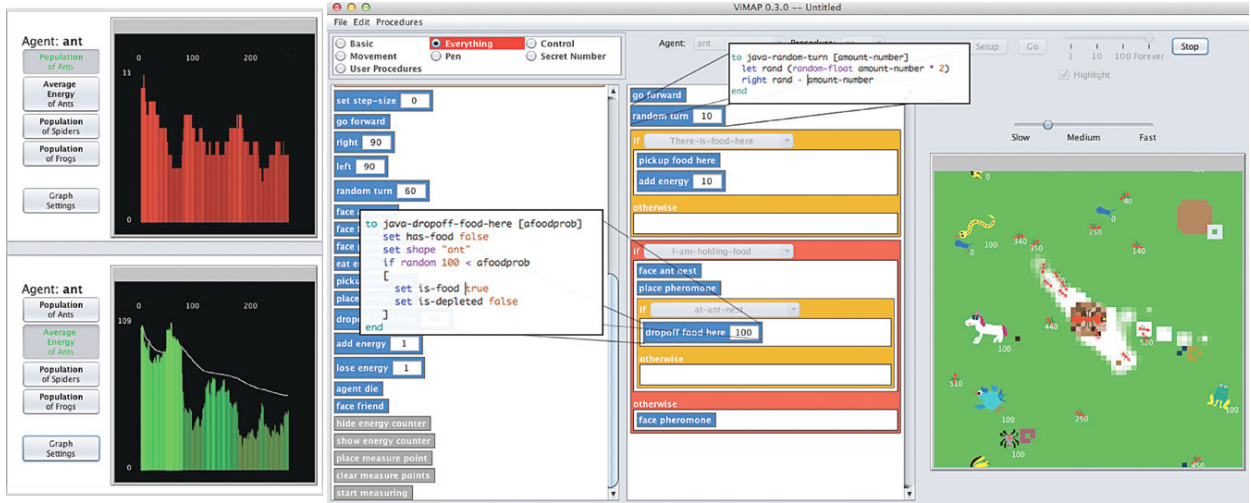

Figure 6: ViMAP-Ant-Foodweb simulation and programming commands developed by 8th-graders. Popped-out images show NetLogo procedures underlying the ViMAP commands created by the students. Left-to-right: Graphs of population and energy; library of ViMAP commands; a sample ViMAP program; and the NetLogo simulation controlled by the ViMAP program.

To address this issue, the teacher then decided to return to the ViMAP Ants unit (see www.vimapk12.net for the curricular activities), and asked the students to work in small groups to create new ViMAP commands by modifying and extending the underlying NetLogo code. For the 8th graders, this work was motivated by a capstone project of designing and creating a version of ViMAP-Ants in order to teach 4th graders about food webs, which is a required curricular standard in 4th grade. The teacher introduced the students to relevant "chunks" (procedures) in the NetLogo code pertaining to specific ViMAP commands they were already familiar with. She led class discussions in which the students collaboratively interpreted and explained the significance of the computational abstractions in NetLogo code in terms of relevant scientific concepts represented in the ViMAP commands. Learning the syntax and new forms of abstractions (such as classes) in text-based programming therefore became deeply intertwined with the relevant concepts in ecology (for example, hierarchy of organisms in food webs). While students' previous work using visual programming introduced them to computational abstractions such as loops, variables, and conditionals, text-based programming further deepened their computational thinking because it involved creating computational objects or classes, declaring new local variables, creating and modifying conditionals, editing and repurposing lists, and using random numbers. Students' growth in computational thinking was further evident in a post-assessment activity, in which they successfully created new commands for a NetLogo simulation of a different ecosystem without teacher-provided assistance (Sengupta et al., 2015b).

### Discussion: Computational Thinking as Experience in K12 STEM

In *Quest for Certainty,* John Dewey argued against empiricist ontology that substitutes data for objects (and inquiry). Data, he argued, signifies a phenomenon for further inquiry; but instead, empiricism often represents data as being self-evident (Dewey, 1929). In a similar vein, the persistent fallacy of the predominant epistemology in educational computing, especially as it pertains to computational thinking in education, is the normative notion that knowledge is some *antecedent reality* (Dewey, 1929), reified in terms of learners' use of computational abstractions used commonly by professional coders. That is, for researchers, the experience of learners is substituted by "computational abstractions" that they used. Certainly, there are efforts, especially by constructionist scholars to demonstrate how computing can take on diverse and personally meaningful forms (Resnick, Berg & Eisenberg, 2004; Farris & Sengupta, 2016), but the hallmark of the experience of coding, as reported in nearly all research articles on computational thinking (including some of our own previous work) remains the deft use of computational abstractions by learners who haven't had much prior experience with coding. This is the danger of technocentrism (Papert, 1987), realized – where the questions about technology are being answered by referring the questions to technology itself.

In this chapter, we have argued for paradigmatic shift in the epistemology and pedagogy of computing and computational thinking, especially for K-12 STEM education. Our position is that we must shift away from empiricist ontology that Dewey argued against (Dewey, 1929) or technocentrism that Papert argued against (Papert, 1987), toward more phenomenological perspectives, both in terms of trying to understand and support the development of computational thinking as *experience* in the context of K12 STEM education. Epistemologically, we have argued that computational thinking must be re-conceptualized more appropriately as an intersubjective experience, as opposed to a more cognitivist image of reasoning that can be assessed through the production of symbolic code. Pedagogically, we have argued that this necessitates careful considerations of the complexities of K12 classrooms, without ignoring teachers' and students' experiences in which computing and coding are situated. In particular, we have proposed the following pedagogical guidelines for sustaining computational thinking in the K12 classroom:

1. Reframing programming and coding as "modeling", in particular, as the design of mathematical units of measurement of change over time, for the K12 science classroom;

2. Trans-disciplinary representational and epistemic practices such as design and modeling can help us support continuity in learning experiences across disciplines;

3. The importance of embodied modeling and non-computational materials as representational and cognitive amplifications of computational code;

4. The role of disciplinarily grounded, normative instructional approaches (e.g., socio-mathematical norms) in refining computational modeling;

5. Reframing coding and modeling as designing for an authentic audience; and

6. The importance of using both visual and text-based programming languages for longer-term curricular integration.

This list is far from exhaustive. However, given the context in which most of our studies have been carried out – high-poverty, predominantly non-white classrooms in public schools with limited resources – we believe that these guidelines can help us focus our attention on issues that can make a difference in terms of helping teachers and students adopt computing and computational thinking as a "language" of STEM, especially on a long-term basis.